\documentclass[lettersize,journal]{IEEEtran}
\usepackage{amsmath,amsfonts}
\usepackage{array}
\usepackage{textcomp}
\usepackage{stfloats}
\usepackage{url}
\usepackage{verbatim}
\usepackage{graphicx}
\usepackage{cite}

\usepackage{subfigure}
\usepackage{array}
\usepackage{multirow}
\usepackage{float}
\usepackage{graphicx}
\usepackage{rotating}
\usepackage{amsmath}
\usepackage{tabularray}
\usepackage{hyperref}
\usepackage{enumerate}

\begin{document}

\title{Auto-scaling Approaches for Microservice Applications: A Survey and Taxonomy}

%\author{\IEEEauthorblockN{Linfeng Wen$^{1, 2}$, Junhan Liao$^{1, 2}$, Minxian Xu$^{1}$, Kejiang Ye$^{1}$, Chengzhong Xu$^{3}$}\\
%\IEEEauthorblockA{1. Shenzhen Institutes of Advanced Technology, 
%Chinese Academy of Sciences, China\\
%2. University of Chinese Academy of Sciences, China\\
%3. State Key Lab of IOTSC, University of Macau, China\\
%\{lf.wen, jh.liao, mx.xu, kj.ye\}@siat.ac.cn, czxu@um.edu.mo}}

\author{Minxian Xu,~\IEEEmembership{Senior Member,~IEEE,}
        Junhan Liao, 
        Linfeng Wen,
        Huaming Wu,~\IEEEmembership{Senior Member,~IEEE,}
        Kejiang Ye,~\IEEEmembership{Senior Member,~IEEE,}
        Rajkumar Buyya,~\IEEEmembership{Fellow,~IEEE}
        Chengzhong Xu,~\IEEEmembership{Fellow,~IEEE}%

\thanks{M. Xu, J. Liao, L. Wen and K. Ye are with Shenzhen Institutes of Advanced Technology, Chinese Academy of Sciences, Shenzhen, China, and University of Chinese Academy of Sciences, Beijing, China.

H. Wu is with Tianjin University, Tianjin, China.

R. Buyya is with the Quantum Cloud Computing and Distributed Systems (qCLOUDS) Laboratory, School of Computing and Information Systems, the University of Melbourne, Melbourne, Australia. 

C. Xu is with State Key Lab of IOTSC, University of Macau, Macau, China.

This work is supported by National Natural Science Foundation of China under Grant 62572462, Guangdong Science and Technology Cooperation Project (No. 2025A0505020065), Guangdong Basic and Applied Basic Research Foundation (No. 2024A1515010251, 2023B1515130002), and Shenzhen Science and Technology Program under Grant JCYJ20240813155810014.

K. Ye is the corresponding author (kj.ye@siat.ac.cn). 
}
}

\markboth{}%
{Shell \MakeLowercase{\textit{et al.}}: A Sample Article Using IEEEtran.cls for IEEE Journals}


\maketitle

\begin{abstract}
Microservice applications are created as loosely coupled application components and they leverage cloud elasticity to reduce costs and increase development speed. However, microservice applications exhibit complex interactions among dynamically evolving services and highly variable workloads, posing significant challenges to auto-scaling mechanisms. Key issues include service dependency management, performance profiling, anomaly detection, workload characterization, and fine-grained resource allocation. To address these challenges, recent auto-scaling approaches leverage historical and runtime data to adapt resource provisioning and optimize system efficiency. Since 2018, marked by the graduation of Kubernetes as the first Cloud Native Computing Foundation (CNCF) project, microservice applications have been widely deployed on standardized orchestration platforms, fundamentally shifting auto-scaling from coarse-grained to service-level, dependency-aware strategies. Accordingly, this paper surveys state-of-the-art auto-scaling approaches for microservice applications since 2018 and presents a taxonomy along five dimensions: infrastructure, architecture, scaling methods, optimization objectives, and behavior modeling. %, Infrastructure defines the environment and constraints for scaling strategies; Behavior Modeling and Scaling Methods analyze system behavior and determine scaling approaches; Application Architecture further clarifies the applicability and implementation of strategies. 
These perspectives collectively target key objectives, including resource efficiency, cost efficiency, and Service Level Agreement (SLA) assurance, aiming to balance system optimization with SLA compliance. We further present a comprehensive comparison and in-depth analysis of representative approaches, examining their core features, strengths, limitations, and applicable scenarios, as well as their performance across diverse environments and workload conditions. %This includes analyzing the trade-offs between different scaling strategies, their impact on resource utilization, system stability, and cost efficiency, as well as how they address specific challenges in cloud-native applications. 
Finally, we synthesize the current research landscape, highlight open challenges and research gaps, and outline promising future directions.%, with particular emphasis on large language model (LLM) driven approaches, microservice dependency–aware scaling, and meta-learning techniques that enhance adaptability and generalization across heterogeneous environments.
\end{abstract}


\section{Introduction}

Microservice-oriented cloud platforms have fundamentally transformed modern computing by offering unprecedented flexibility, scalability, and cost efficiency to enterprises and organizations worldwide \cite{I1, I2, I3}. By decomposing applications into loosely coupled, independently deployable services, this architectural paradigm relieves organizations from managing complex monolithic infrastructures and instead allows them to focus on core business innovation and faster time-to-market for new products and services. To support this paradigm, major cloud service providers, including Amazon, Google, and Alibaba, have established large-scale cloud platforms that deliver robust infrastructure and elastic services, enabling microservice applications to scale efficiently across organizations of varying sizes \cite{I9}.

With the advancement of cloud-native technologies, the adoption of containerization and microservice architectures has increased significantly \cite{I10}. By encapsulating individual microservices into isolated containers, these technologies simplify deployment and operations while improving system reliability and maintainability \cite{I11}. The emergence of container orchestration platforms such as Kubernetes \cite{I12} has further enabled effective management and scheduling of large-scale microservice applications. To cope with dynamic workloads, Kubernetes provides built-in auto-scaling mechanisms that adjust application resources according to real-time demand \cite{I14}. Specifically, Kubernetes supports two primary auto-scaling strategies: the Horizontal Pod Autoscaler (HPA) and the Vertical Pod Autoscaler (VPA). The HPA dynamically scales the number of pod replicas based on CPU utilization or custom metrics, allowing microservices to respond to workload fluctuations, while the VPA improves resource efficiency by adjusting pod resource requests and limits to better match actual consumption. Together, these mechanisms offer a baseline resource management solution for microservice applications, enabling Kubernetes users to handle evolving workloads in a largely automated manner.

%\subsection{The Need for Advanced Auto-scaling Technology}

However, as microservice applications continue to evolve and application scenarios become increasingly diverse, the requirements for auto-scaling have grown more stringent and heterogeneous \cite{Related1}. Traditional auto-scaling mechanisms, which primarily rely on reactive adjustments based on coarse-grained metrics, are often insufficient for handling the complexity and dynamics of real-world microservice deployments. These practical limitations motivate the need to address the following key challenges:

\subsubsection{The need for more intelligent and accurate auto-scaling strategies}
Traditional heuristic-based auto-scaling approaches often rely on predefined thresholds and reactive policies, which makes them inadequate for anticipating rapid or irregular workload fluctuations in microservice applications. As a result, more intelligent strategies that incorporate machine learning and data-driven analysis are increasingly required to capture workload dynamics, predict demand trends, and enable timely and precise scaling decisions.

\subsubsection{The demand to manage resource contention and inter-application performance interference in co-located environments}
Existing auto-scaling approaches often struggle to effectively manage resource contention and performance interference in shared microservice environments \cite{I24, I25}. As multiple microservices compete for underlying resources, performance degradation, increased operational overhead, and potential SLA violations may arise.

\subsubsection{The requirement to incorporate for inter-dependencies and invocation relationships among microservices}
Microservice applications exhibit complex inter-service dependencies and invocation relationships that have a direct impact on scaling behavior and end-to-end performance \cite{I26}. Incorporating dependency management and invocation-path–aware analysis into auto-scaling decisions enables more coordinated scaling actions, helping to avoid performance bottlenecks, mitigate fault propagation, and ultimately improve system scalability and reliability.

\subsubsection{The need for more comprehensive and refined metrics and monitoring}
Conventional auto-scaling mechanisms, such as the HPA and VPA, primarily rely on coarse-grained resource metrics, including CPU and memory utilization. However, in microservice applications, additional factors, such as network bandwidth, disk I/O, and service-level performance indicators, are critical for accurately capturing application health and making informed scaling decisions \cite{I29, I30, I31}.

%\subsubsection{The necessity to consider the time cost of auto-scaling}
%The time cost of auto-scaling includes the detection time of load changes, decision-making time, and the time required to execute scaling strategies. Long scaling time can affect performance and availability, making it essential to balance time costs with scaling efficiency.

%\subsection{Motivations}

 %As shown in the Cloud-native Computing Foundation survey report of 2023\footnote{https://www.cncf.io/reports/cncf-annual-survey-2023/} in Fig.~\ref{fig:1}, 
The widespread adoption of microservice-based cloud platforms has brought auto-scaling to the forefront of modern IT systems. While Kubernetes has become the dominant platform for auto-scaling in practice, alternative solutions continue to emerge, and significant opportunities remain for research, development, and real-world deployment of advanced auto-scaling techniques. As technologies and application requirements evolve, auto-scaling mechanisms are expected to deliver increasingly efficient, flexible, and intelligent resource management for microservice applications, driving progress across the cloud systems ecosystem. Accordingly, this survey aims to investigate and synthesize recent advances in auto-scaling approaches for microservice-oriented architectures. The key \textbf{contributions} of this paper are summarized as follows:

\begin{itemize} 
    
    \item [$ \bullet $] We present a comprehensive and up-to-date survey of auto-scaling research, summarizing the key contributions of existing studies and systematically comparing and analyzing representative works.
    
    \item [$ \bullet $] We review recent auto-scaling approaches for microservice applications, examining their core characteristics, limitations, and applicability, and highlighting the methodological evolution of auto-scaling techniques in recent years.
    
    \item [$ \bullet $] We propose an extensive taxonomy for microservice auto-scaling, covering the majority of existing approaches and categorizing them according to their key design features and operating conditions, followed by detailed comparative analysis.
    
    \item [$ \bullet $] We identify open challenges and emerging opportunities in auto-scaling for microservice-based systems, and outline promising future research directions by synthesizing insights from current literature.
    
\end{itemize} 

\begin{table*}
\centering
\caption{Comparison of Related Survey and Taxonomy Works.}
\label{Comparison}
\resizebox{\linewidth}{!}{%
\begin{tblr}{
  cells = {c},
  cell{1}{1} = {r=2}{},
  cell{1}{2} = {r=2}{},
  cell{1}{3} = {r=2}{},
  cell{1}{4} = {c=7}{},
  cell{1}{11} = {r=2}{},
  hline{1,9} = {-}{0.08em},
  hline{3} = {-}{},
}
\textbf{Reference} & \textbf{Type} & {\textbf{Year}} & \textbf{Taxonomy} &  &  &  &  &  &  & \textbf{Focus}\\
 &  &  & \textbf{Architecture} & \textbf{Modeling} & \textbf{Objective} & \textbf{Technique} & \textbf{Metrics} & \textbf{Infrastructure} & \textbf{Method} & \\
 \cite{Related1} & Survey & 2018 & \checkmark & \texttimes & \texttimes & \checkmark & \checkmark & \texttimes & \checkmark & Web Apps\\
 \cite{Related2} & Survey & 2018 & \texttimes & \texttimes & \texttimes & \checkmark & \checkmark & \checkmark & \texttimes & Traditional Cloud Apps\\
 \cite{Related3} & Survey & 2023 & \checkmark & \texttimes & \texttimes & \checkmark & \checkmark & \checkmark & \checkmark & Hybrid Fog/Edge/Cloud Apps\\
 \cite{Related5} & Review & 2021 & \texttimes & \checkmark & \texttimes & \checkmark & \checkmark & \checkmark & \checkmark & IoT-based apps, VM \\
 \cite{Related6} & Review & 2021 & \texttimes & \texttimes & \checkmark & \checkmark & \texttimes & \checkmark & \checkmark & Traditional Cloud Apps\\
This work & Survey & 2026 & \checkmark & \checkmark & \checkmark & \checkmark & \checkmark & \checkmark & \checkmark & Microservice Apps
\end{tblr}
}
\end{table*}

\section{Related Work}

As the scale and complexity of microservice-based applications continue to grow, auto-scaling has become an increasingly active research area. Several survey and review studies have examined cloud auto-scaling from different perspectives. This section provides a comparative analysis of representative auto-scaling surveys closely related to our work. To identify relevant studies, we conducted a systematic search on Google Scholar, focusing on publications from the past five years, limited to survey or review articles published in SCI-indexed journals included in JCR. Based on these criteria, five closely related works were selected for comparison.

From a technical perspective, Qu et al. \cite{Related1} presented a comprehensive survey and taxonomy of auto-scaling mechanisms for cloud-based web applications, focusing on system challenges and existing solutions. However, their study primarily targets monolithic application architectures and reactive scaling strategies, which limits its applicability to modern microservice environments characterized by fine-grained services and complex dependencies. To address broader system adaptability, Chen et al. \cite{Related2} surveyed adaptive cloud auto-scaling systems that respond to runtime changes across heterogeneous hardware and configurations to support cloud elasticity. Nevertheless, the approaches discussed largely focus on traditional cloud infrastructures, and do not fully reflect recent advances in auto-scaling for microservice applications, nor the emergence of more powerful learning-based techniques, such as Transformer-based models.%Saif et al. \cite{Related6}, in their large-scale study on microservice invocation dependencies at Alibaba, found interesting results that will significantly impact microservice auto-scaling technologies.

From the perspective of taxonomy and analytical depth, Dogani et al. \cite{Related3} surveyed auto-scaling techniques in containerized cloud computing as well as in edge and fog environments, and outlined potential future research directions. While their study provides a broad technical overview, it focuses primarily on scaling mechanisms themselves, without systematically classifying infrastructure layers, optimization objectives, or resource provisioning strategies. Moreover, it does not explicitly address microservice dependency structures or resource contention among co-located applications, which are central challenges in microservice auto-scaling.
Verma and Bala \cite{Related5} examined contemporary auto-scaling techniques together with load forecasting and virtual machine migration, presenting a technical taxonomy and qualitative assessment. However, their survey mainly targets IoT-oriented cloud applications and does not consider auto-scaling in microservice-based architectures. In addition, it lacks a structured classification across infrastructure, scaling objectives, and resource provisioning dimensions.
%Saif et al. \cite{Related6} reviewed autonomous resource management approaches and categorized them according to design principles, objectives, functional components, and application domains, providing qualitative comparisons of their strengths and limitations. Nevertheless, this work addresses autonomous resource management in a broad sense and does not offer a comprehensive or focused analysis of auto-scaling technologies, particularly in the context of microservice applications.

As summarized in Table~\ref{Comparison}, we present a detailed comparison with existing survey and review works. Our survey focuses on high-quality research published over the past five years, with particular emphasis on recent innovations in auto-scaling and their applicability to microservice applications. Unlike monolithic architectures, microservice systems consist of multiple loosely coupled services, which introduces unique challenges related to inter-service invocations, dependency management, and resource contention and performance interference among co-located applications—issues that are largely overlooked in prior surveys. Building upon existing classification schemes, we further propose a refined and comprehensive taxonomy that systematically captures the design space of auto-scaling techniques and accommodates the distinguishing characteristics of microservice-based systems, thereby providing a structured foundation and guidance for future research in this domain.

%\subsection{Structure}

%The subsequent sections of the paper are structured as follows: Section~\ref{2} presents an overview of cloud-native architecture and auto-scaling technologies. Section~\ref{3} provides a comprehensive examination of the taxonomy of auto-scaling technologies for cloud-native applications. Section~\ref{4} conducts a taxonomy study of the surveyed papers and analyzes the recent evolution of auto-scaling technologies. Section~\ref{5} provides a conclusion and future direction. Lastly, Section ~\ref{6} concludes the findings and discussions presented in this paper.

\section{Background}\label{2}

In this section, we provide an overview of microservice-oriented cloud architectures and the associated auto-scaling technologies.

\begin{figure*}
	\centering
	\subfigure[Containerized Deployment.]{
		\includegraphics[width=0.28\textwidth]{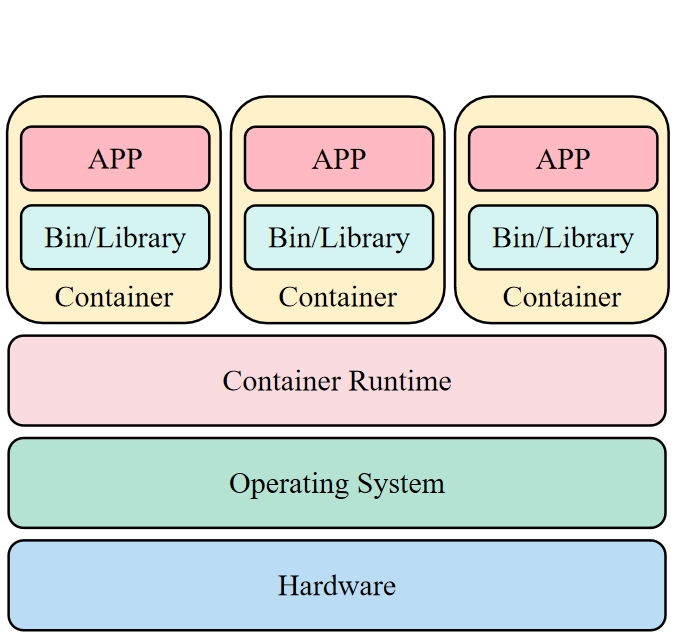}}
			\subfigure[Architecture of the Train-Ticket.]{
		\includegraphics[width=0.49\textwidth]{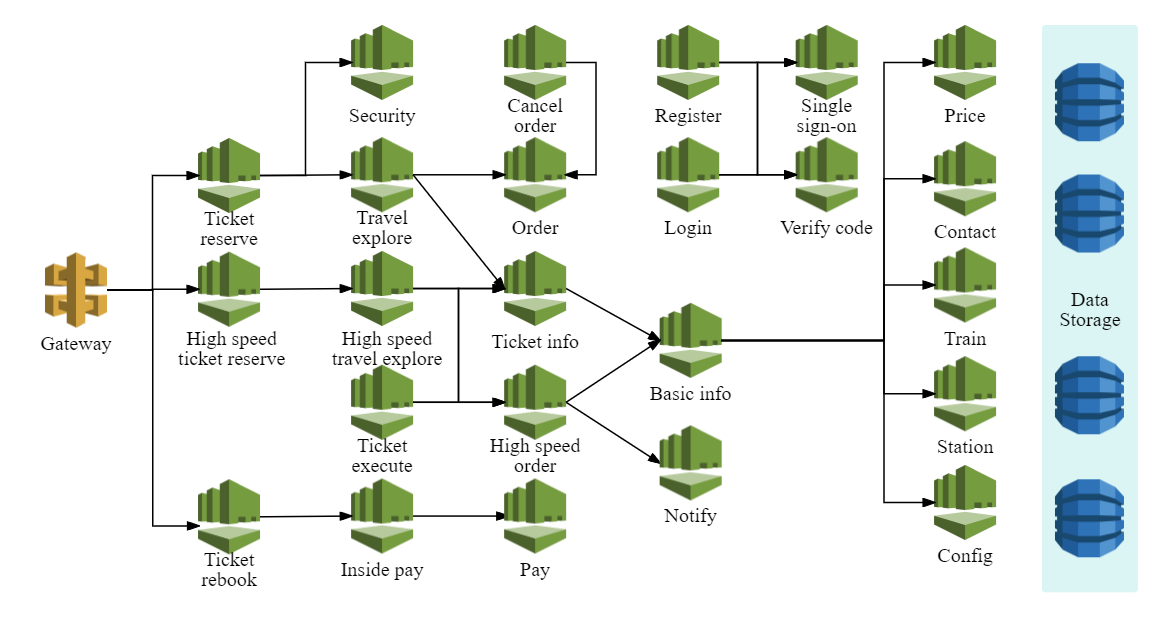}}
	\caption{Microservice Application Deployment Architecture and A Typical Application Example.}
	\label{fig:7A}
\end{figure*}

%\begin{figure*}
%	\centering
%	\subfigure[Traditional Deployment (1960 - 2000).]{
%	    \includegraphics[width=0.32\textwidth]{fig3-a.png}}
%	\subfigure[Virtualized Deployment (2000 - 2013).]{
%		\includegraphics[width=0.32\textwidth]{fig3-b.png}}
%	\subfigure[Containerized Deployment (2013 - present).]{
%		\includegraphics[width=0.32\textwidth]{fig3-c.png}}
%	\caption{Three Phases of Application Deployment Architecture.}
%	\label{fig:7A}
%\end{figure*}
%\begin{figure*}
%	\centering
% 	\subfigure[Architecture of the Sock-Shop\protect\footnotemark.]{\label{fig:6}
% 	    \includegraphics[width=0.472\textwidth]{fig4-a.png}}
%	\subfigure[Architecture of the Hotel-Reservation \cite{I35}.]{
%		\includegraphics[width=0.437\textwidth]{fig4-b.png}}
%	\subfigure[Architecture of the Train-Ticket \cite{I36}.]{
%		\includegraphics[width=0.49\textwidth]{fig4-c.png}}
%	\caption{Architectures of Two Cloud-native Applications.}
%	\label{fig:7B}
%\end{figure*}

\subsection{Microservice-Oriented Architecture}

Application deployment architectures have evolved through three major stages: traditional, virtualized, and containerized (cloud-native) deployment models \cite{I38}. As illustrated in Fig.~\ref{fig:7A}(a), containerized deployment enables microservice applications to be packaged into lightweight containers using technologies such as Docker \cite{I40}. These containers share the host operating system kernel, resulting in faster startup times, improved resource efficiency, and greater portability, while facilitating fine-grained scaling and seamless migration across heterogeneous environments. %the traditional deployment model \cite{I41}, applications run directly on the operating system of physical hardware. This method is simple but lacks flexibility and resource isolation, making the system vulnerable to stability issues. In virtualized deployment \cite{I39}, a "Hypervisor" allows multiple VMs to run independently on the same hardware, offering better isolation and security but at the cost of higher resource overhead. 

As illustrated in Fig.~\ref{fig:7A}(b), a typical cloud-native application architecture based on microservice design is shown. This architecture decomposes monolithic applications into a collection of independent, loosely coupled microservices, each responsible for a specific business function and interacting with others through well-defined interfaces. Such loose coupling enhances system flexibility, scalability, and evolution agility, making microservice architectures well suited for dynamic and large-scale application environments.

\footnotetext{https://github.com/microservices-demo/microservices-demo}

\subsection{Auto-scaling Technology}

Auto-scaling automatically adjusts resource allocations in response to real-time workload variations, allowing microservice applications to maintain performance and stability while improving resource utilization and cost efficiency. By elastically provisioning and releasing resources, such as scaling service instances during traffic spikes and downsizing during idle periods, auto-scaling reduces operational overhead, energy consumption, and infrastructure costs. Most auto-scaling mechanisms follow the MAPE loop \cite{I47}, comprising monitoring, analysis, planning, and execution stages, which operate continuously to enable adaptive and efficient resource management.
% \begin{figure}
% 	\centerline{\includegraphics[width=\linewidth]{fig6.png}}
% 	\caption{The Architecture of Kubernetes.}\label{fig:25}
% \end{figure}

%\begin{itemize} 
    
%\item [$ \bullet $] \textbf{Monitoring:} The system collects performance metrics, logs, events, and alerts to understand the current status and patterns. For example, Prometheus\footnote{https://prometheus.io/} collects data, Grafana\footnote{https://grafana.com/} visualizes it, and Jaeger\footnote{https://www.jaegertracing.io/} helps with transaction debugging in microservices.

%\item [$ \bullet $] \textbf{Analyzing:} The system analyzes collected data to detect trends, patterns, and anomalies, using predefined rules or models. For example, reinforcement learning \cite{I19} and workload prediction models \cite{xuEsDNNDeepNeural2022} are used for this analysis.

%\item [$ \bullet $] \textbf{Planning:} Based on the analysis, the system creates plans to adjust resources, including scaling strategies and when to take action.

%\item [$ \bullet $] \textbf{Executing:} The system carries out the scaling actions, such as adding VM instances, increasing containers, or adjusting load balancing, through specific API calls.

%\end{itemize} 

% \begin{figure*}
% 	\centerline{\includegraphics[width=0.95\linewidth]{fig5.png}}
% 	\caption{The MAPE Loop of Auto-scaling.}\label{fig:24}
% \end{figure*}

To manage large-scale containerized microservice environments and enable auto-scaling, enterprises widely adopt container orchestration platforms. Popular systems include Kubernetes\footnote{https://kubernetes.io/}, Docker Swarm\footnote{https://docs.docker.com/reference/cli/docker/swarm/}, and Apache Mesos\footnote{https://mesos.apache.org/}. Among these, Kubernetes has emerged as the de facto industry standard, and consequently, most existing auto-scaling research and practice are built upon it. Nevertheless, in large-scale and multi-tenant microservice environments, traditional auto-scaling approaches often struggle to cope with highly dynamic workloads, complex service dependencies, and resource contention, motivating the need for more advanced and intelligent auto-scaling solutions.

\section{A Taxonomy of Auto-Scaling Technologies for Microservice Applications}\label{3}

\begin{figure*}
	\centerline{\includegraphics[width=0.8\linewidth]{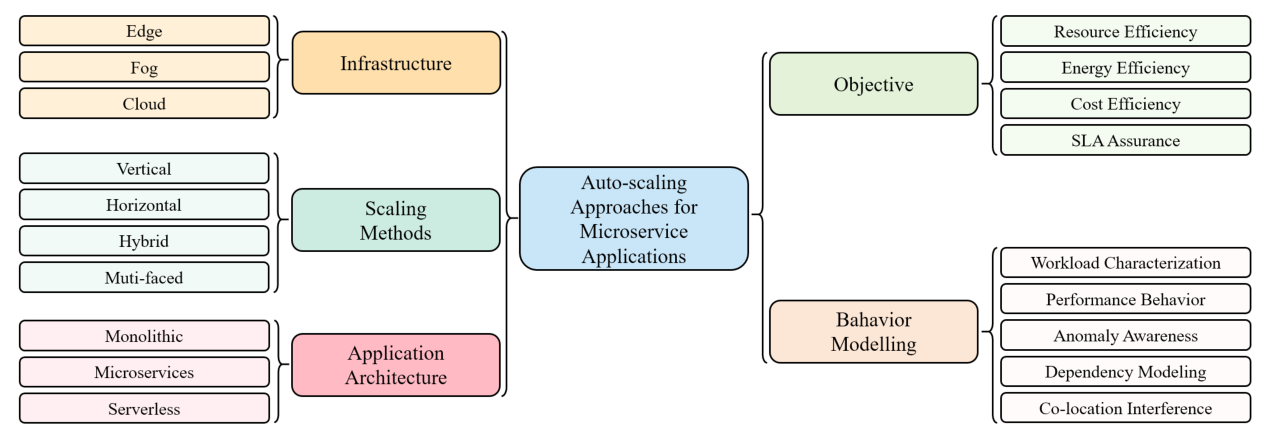}}
	\caption{Taxonomy of Auto-scaling Approaches for Microservice Applications.}\label{fig:12}
\end{figure*}

As illustrated in Fig.~\ref{fig:12}, our taxonomy organizes auto-scaling technologies along five key dimensions, each reflecting a fundamental design decision in microservice auto-scaling systems. \textit{Infrastructure} captures the execution environment and resource abstraction layer on which auto-scaling operates, as different platforms (e.g., cloud, edge, or hybrid environments) impose distinct constraints and capabilities. \textit{Application Architecture} characterizes how applications are structured at the software level, since architectural choices, such as monolithic versus microservice designs, directly affect the granularity and coordination of scaling actions. \textit{Scaling Methods} describe the mechanisms and strategies used to adapt resources to workload dynamics, representing the core operational logic of auto-scaling. \textit{Objectives} define the optimization goals that drive scaling decisions, including performance, efficiency, and reliability, which often involve trade-offs. Finally, \textit{Behavior Modeling} addresses how system and workload dynamics are captured and predicted, as accurate modeling is essential for proactive and adaptive scaling in complex environments.

Together, these dimensions provide a systematic and orthogonal view of the auto-scaling design space, enabling a comprehensive analysis of existing approaches while balancing resource efficiency, cost efficiency, and SLA assurance, and reconciling system optimization with SLA fulfillment.

%This hierarchical structure systematizes microservice auto-scaling strategies from the foundation to the ultimate objectives: Infrastructure (1st layer) defines the environment and constraints for scaling strategies; Behavior Modeling and Scaling Methods (2nd layer) analyze system behavior and determine scaling approaches; Application Architecture (3rd layer) further clarifies the applicability and implementation of strategies. These layers ultimately serve objectives such as resource efficiency, cost efficiency, and SLA assurance, achieving a balance between optimization and SLA fulfillment.

\subsection{Infrastructure}

Auto-scaling technologies are designed for different deployment environments, including edge, fog, and cloud computing, each addressing distinct requirements and constraints. Together, these paradigms form a multi-layered infrastructure continuum that supports microservice applications with diverse latency, scalability, and resource demands. As a result, auto-scaling strategies vary across infrastructures, trading off responsiveness, scalability, and global coordination based on resource availability and deployment scope.

\begin{itemize} 
    
    \item [$ \bullet $] \textbf{Edge Computing:} Brings computation and data processing closer to end users and data sources, typically at network edges or devices \cite{T1}. Auto-scaling at the edge focuses on dynamically adjusting limited computing and storage resources to accommodate workload variations while minimizing latency and improving responsiveness for microservice-based services.

    \item [$ \bullet $] \textbf{Fog Computing:} Acts as an intermediate layer between edge and cloud environments \cite{T4}. It mitigates the resource constraints of edge nodes by coordinating resource allocation across edge and cloud layers, enabling demand-aware auto-scaling that balances performance and resource efficiency.
    
    \item [$ \bullet $] \textbf{Cloud Computing:} Delivers elastic computing resources over the internet, allowing applications to provision and release resources on demand without owning physical infrastructure \cite{T3}. Cloud-based auto-scaling focuses on large-scale elasticity and global optimization to support microservice deployments under highly dynamic workloads.

\end{itemize}

\subsection{Application Architecture}

Auto-scaling techniques are closely tied to the application architecture they support, as architectural choices determine scaling granularity, control scope, and system complexity. While this survey primarily focuses on microservice-based cloud-native applications, existing auto-scaling research also includes relevant work targeting monolithic and serverless architectures. We therefore categorize the literature along these three architectural styles.

\begin{itemize} 

\item [$ \bullet $]\textbf{Monolithic Architecture:} 
Monolithic applications are deployed as a single, tightly coupled unit. Auto-scaling in this setting is typically coarse-grained, adjusting resources at the application or virtual machine level based on aggregated workload metrics. Several early auto-scaling studies adopt this model due to its simplicity, making monolithic systems an important baseline for understanding foundational scaling strategies.

\item [$ \bullet $]\textbf{Microservices Architecture:} 
Microservices decompose applications into independently deployable services, enabling fine-grained, service-level scaling. This architecture improves elasticity and resource efficiency but introduces challenges such as inter-service dependencies, load propagation, and shared resource contention. As these challenges motivate a large body of recent work, microservices constitute the primary architectural focus of this survey.

\item [$ \bullet $]\textbf{Serverless Architecture:} 
Serverless architectures execute applications as event-driven functions managed by cloud providers \cite{T5}. Scaling is largely handled automatically by the platform, abstracting infrastructure management from developers. Related research often addresses issues such as cold-start latency, cost efficiency, and provider-level scheduling, which differ from the explicit control-oriented auto-scaling mechanisms studied in microservice-based systems.

\end{itemize}

\subsection{Scaling Methods}

Scaling methods determine how resources are adjusted under workload dynamics and are closely coupled with application architecture, as different architectures impose distinct constraints on scaling granularity, responsiveness, and coordination overhead. Based on existing literature, we categorize auto-scaling methods into four representative types.

\begin{itemize} 
    
\item [$ \bullet $]\textbf{Vertical Scaling:} 
Increases the resource capacity of individual instances (e.g., CPU and memory), providing fast performance improvement but limited by hardware constraints and reconfiguration overhead.

\item [$ \bullet $]\textbf{Horizontal Scaling:} 
Adds or removes instances such as virtual machines or containers, offering high elasticity and fault tolerance and serving as the dominant approach in microservice-based systems.

\item [$ \bullet $]\textbf{Hybrid Scaling:} 
Jointly applies vertical and horizontal scaling to balance capacity expansion and parallelism, enabling flexible adaptation to diverse and highly dynamic workloads.

\item [$ \bullet $]\textbf{Multi-faceted Scaling:} 
Augments hybrid scaling with complementary mechanisms, such as service-level adaptation or workload degradation (e.g., brownout \cite{brownout}), to improve robustness and resource efficiency under extreme load conditions.

\end{itemize}

\subsection{Scaling Objectives}

Scaling objectives define the optimization goals that guide auto-scaling decisions and directly influence strategy design under varying workloads and operational constraints. Existing studies commonly focus on the following objectives.

\begin{itemize} 
    
\item [$ \bullet $] \textbf{Resource Efficiency:} 
Aligns allocated resources (e.g., CPU, memory, instances) with workload demand to maximize utilization and avoid over- or under-provisioning.

\item [$ \bullet $] \textbf{Energy Efficiency:} 
Reduces power consumption by consolidating workloads and scaling down idle resources, which is particularly important in large-scale data centers.

\item [$ \bullet $] \textbf{Cost Efficiency:} 
Minimizes operational expenses in pay-as-you-go cloud environments by provisioning resources only when needed.

\item [$ \bullet $] \textbf{SLA Assurance:} 
Ensures application performance, availability, and response latency under workload fluctuations, preventing quality of service (QoS) degradation and SLA violations.

\end{itemize}

\subsection{Behavior Modeling}

Behavior modeling captures how workloads, applications, and system components interact over time and under changing conditions, serving as the analytical foundation for informed auto-scaling decisions. By modeling system behavior beyond instantaneous metrics, auto-scaling mechanisms can anticipate demand changes, mitigate interference, and coordinate scaling actions more effectively. Existing studies primarily focus on the following aspects.

\begin{itemize} 
    
\item [$ \bullet $] \textbf{Workload Characterization:} 
Models workload patterns, trends, and bursts to improve demand prediction and proactive scaling.

\item [$ \bullet $] \textbf{Performance Behavior:} 
Analyzes response latency, throughput, and resource utilization to ensure scaling actions preserve performance objectives.

\item [$ \bullet $] \textbf{Anomaly Awareness:} 
Detects abnormal behaviors such as sudden load spikes or performance degradation, enabling timely corrective actions.

\item [$ \bullet $] \textbf{Dependency Modeling:} 
Captures service invocation and dependency relationships to avoid cascading bottlenecks during scaling.

\item [$ \bullet $] \textbf{Co-location Interference:} 
Accounts for resource contention and performance interference among co-located tasks or services in shared environments.

\end{itemize}

\subsection{Evolution of Auto-scaling Approaches for Microservice Applications}

\begin{figure}
	\centerline{\includegraphics[width=\linewidth]{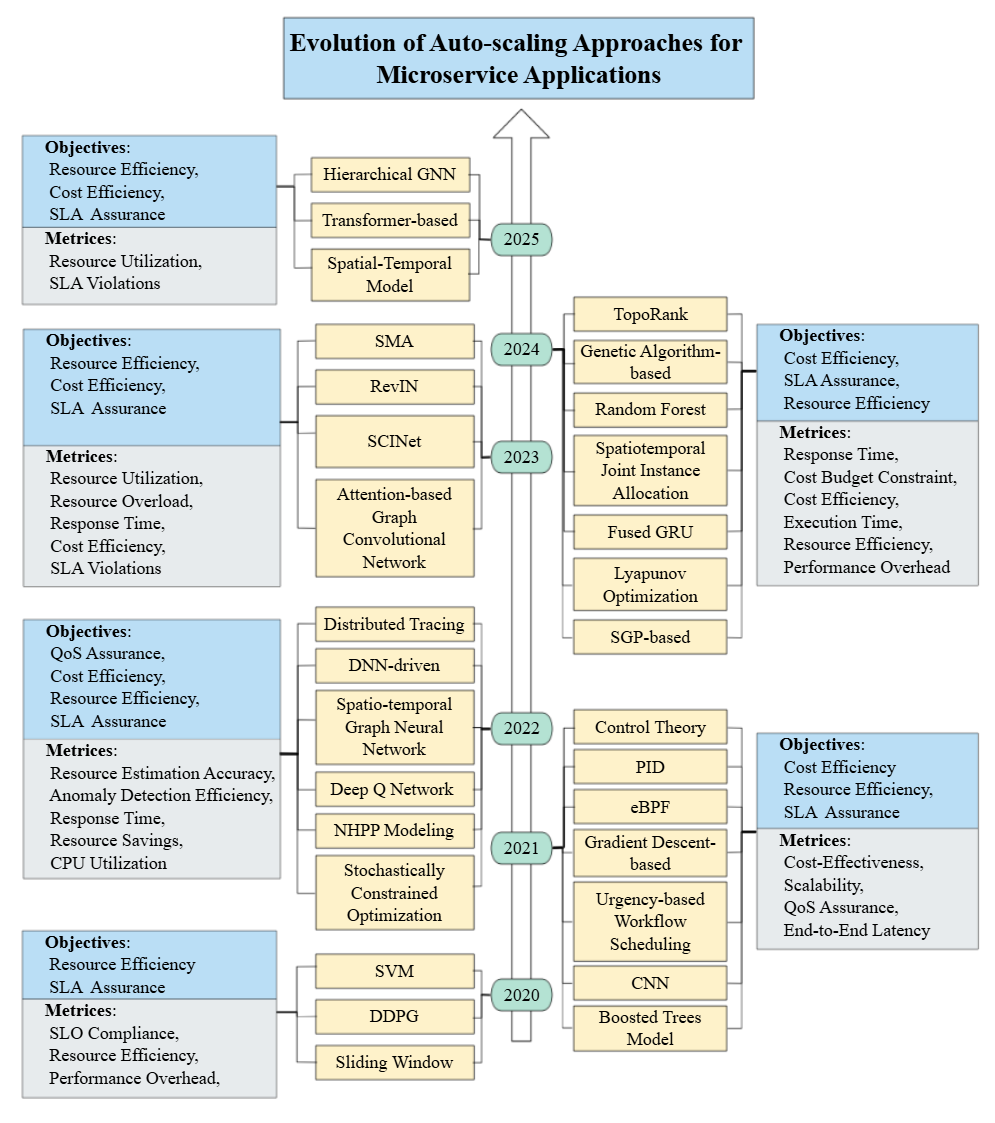}}
	\caption{Evolution of Auto-scaling Approaches for Microservice Applications.}\label{fig:Evolution}
\end{figure}

Auto-scaling techniques have evolved with the increasing complexity of microservice applications. Early rule-based methods using coarse-grained metrics proved insufficient for fine-grained service management. Recent research focuses on per-microservice scaling, dependency awareness, and predictive or learning-based coordination to improve resource efficiency and ensure end-to-end performance and SLA compliance, as shown in Fig.~\ref{fig:Evolution}.

Recent research on auto-scaling for microservice applications has progressively moved from reactive, resource-centric mechanisms toward predictive, learning-driven, and dependency-aware solutions.

Early work in 2020 and before emphasized mitigating SLA violations and resource contention in shared microservice cluster. Qiu et al. \cite{I19} combined support vector machine (SVM)-based SLA violation detection with reinforcement learning (RL) to reduce contention on critical execution paths, while Google’s Autopilot system \cite{rzadcaAutopilotWorkloadAutoscaling2020} demonstrated the effectiveness of joint horizontal–vertical scaling through data-driven policies for production workloads.

In 2021, auto-scaling research increasingly incorporated control-theoretic and optimization-based techniques to improve per-microservice elasticity. Studies explored Proportional-Integral-Derivative (PID) based controllers \cite{baarziSHOWARRightSizingEfficient2021}, gradient-based SLA allocation \cite{mirhosseiniParsloGradientDescentbased2021}, and cost-aware scheduling models \cite{wangElasticSchedulingMicroservice2021}. Learning-based methods also emerged to capture workload dynamics and QoS constraints \cite{zhangSinanMLbasedQoSaware2021}.

By 2022, predictive modeling became central, with deep learning and graph-based methods used to forecast workload and resource demand while explicitly considering service interactions. Representative works employed deep neural networks (DNNs) \cite{chowDeepRestDeepResource2022}, spatiotemporal graph neural networks (GNNs) combined with reinforcement learning \cite{wangDeepScalingMicroservicesautoscaling2022}, and stochastic optimization frameworks to balance cost and QoS \cite{qianRobustScalerQoSAwareAutoscaling2022}.

Research in 2023 further advanced proactive and dependency-aware auto-scaling. Studies leveraged time-series forecasting models \cite{jeongProactiveResourceAutoscaling2023a}, online prediction mechanisms \cite{chengProScaleProactiveAutoscaling2023}, and graph learning to uncover hidden microservice dependencies and improve coordinated scaling decisions \cite{mengDeepScalerHolisticAutoscaling2023a}.

In 2024, auto-scaling approaches have focused on bottleneck awareness, system-wide optimization, and emerging execution models. Works introduced topology-aware bottleneck detection for microservices \cite{xiePBScalerBottleneckAwareautoscaling2024}, budget-constrained QoS optimization \cite{chengGeoScaleMicroserviceAutoscaling2024}, and proactive elasticity for serverless environments \cite{fengHeterogeneityawareProactiveElastic2024}, reflecting a trend toward holistic and cost-aware scaling strategies.

Most recently, in 2025, auto-scaling research for microservice applications has advanced toward predictive, dependency-aware, and resource-efficient strategies. Works introduced a multi-layered framework combining predictive scaling and service migration across containers and VMs to reduce service level objective (SLO) violations and infrastructure costs \cite{SHAFI2025104266}; a spatio-temporal GNN with deep reinforcement learning (DRL) approach that models microservice dependencies as dense connectivity graphs, improving resource prediction accuracy and reducing SLA violations \cite{LIANG2026107909}; and a reactive and proactive horizontal pod auto-scalers that optimize Kubernetes scaling in resource-constrained environments, mitigating over- and under-provisioning \cite{AHMAD2025112390}. These efforts reflect a trend toward holistic, workload-aware, and cost-sensitive auto-scaling solutions tailored to complex microservice applications.

Overall, a wide range of techniques has been applied to the auto-scaling of microservice applications, including workload modeling, performance analysis, anomaly detection, dependency/link analysis, and interference management. These methods have gradually evolved from basic threshold- and heuristic-based approaches, as well as control- and queuing-theory methods, to more intelligent and adaptive solutions leveraging machine learning, deep learning, and reinforcement learning. To handle complex microservice chains and highly dynamic workloads, advanced models such as graph neural networks and attention mechanisms have been adopted, enhancing prediction accuracy and resource allocation efficiency, which are essential for maintaining the stability, performance, and SLA compliance of microservice applications.

\section{State-of-the-Art in Auto-scaling for Microservice Applications}\label{4}

In this section, we provide a comprehensive review of auto-scaling approaches for microservice applications. To highlight the key contributions of the reviewed studies, we adopt the taxonomy presented in Section~\ref{3}, with a primary focus on behavior modeling, as understanding workloads, performance dynamics, service dependencies, and interference patterns is critical for effective and reliable auto-scaling in complex microservice environments. Comparisons of infrastructure, application architecture, scaling methods, and scaling objectives are interpreted through the lens of behavior modeling, emphasizing how each design choice impacts workload prediction, resource allocation, and SLA compliance. While some works span multiple categories, each study is classified according to the primary research challenge it addresses, allowing a systematic analysis of how behavior-oriented techniques drive improvements in microservice auto-scaling.

\subsection{Article Selection Methodology}

In this section, we describe our methodology for identifying relevant literature on microservice auto-scaling. We performed a comprehensive search across leading academic databases, including ACM Digital Library, IEEE Xplore, Springer, Elsevier, Usenix, ScienceDirect, Wiley Interscience, and Google Scholar. Search queries targeted titles and abstracts using keywords such as auto-scaling, microservice, serverless, elastic scaling, dynamic scaling, efficient scheduling, proactive scaling, resource management, and resource efficiency.

Given the large number of initial search results, many of which were only tangentially related, we applied secondary screening based on relevance, focusing on studies that specifically addressed auto-scaling challenges, strategies, or evaluation in microservice environments. To ensure the quality and reliability of the selected literature, only SCI- and EI-indexed publications were included. This approach guarantees that the survey reflects well-established, peer-reviewed research while minimizing bias from less rigorous or anecdotal studies.

Through this process, we curated 50 research articles (24 conference papers and 26 journal articles), all directly addressing microservice auto-scaling. Each work was assessed based on the relevance and rigor of its methodology, the practicality of scaling, performance in complex microservice deployments, as well as novelty and effectiveness of scaling techniques. This selection process ensures the survey focuses on the most impactful and credible contributions in the field.

\subsection{Behavior Modeling}

\subsubsection{\textbf{Workload Characterization}}

\begin{table*}
\centering
\caption{Classification Based on Workload Characterization.}
\label{Workload Characterization}
\resizebox{\linewidth}{!}{%
\begin{tblr}{
  cells = {c},
  hline{1,2,15} = {-}{0.08em},
  row{1-Z} = {m, 1cm},
  rowsep = 0.5em  % Adjust this value to reduce row spacing
}
\textbf{Reference} & \textbf{Architecture} & \textbf{Infrastructure} & {\textbf{Scaling}\\\textbf{Method}} & \textbf{Technique} & {\textbf{Scaling}\\\textbf{Indicator}} & {\textbf{Scaling}\\\textbf{Timing}} & \textbf{Objective} & \textbf{Year} & \textbf{Source}\\ 
 \cite{wenTempoScaleCloudWorkloads2024} & Microservices & Cloud & Vertical & {Transformer-based\\Deep Learning} & System Metrics & Proactive & {Resource Efficiency\\SLA Assurance} & 2024 & IEEE CLOUD\\
 \cite{xuCoScalMultifacetedScaling2022} & Microservices & Cloud & Muti-faceted & {Reinforcement Learning\\Deep Learning} & Hybrid & Proactive & SLA Assurance & 2022 & IEEE TNSM\\
 \cite{jeongProactiveResourceAutoscaling2023a} & Microservices & Cloud & Hybrid & {Deep Learning\\Threshold-based} & System Metrics & Hybrid & {Resource Efficiency\\Cost Efficiency} & 2023 & IEEE TCC\\
 \cite{qianRobustScalerQoSAwareAutoscaling2022} & Microservices & Cloud & Horizontal & {Control Theory\\Queuing Theory} & Business Metrics & Proactive & {SLA Assurance\\Cost Efficiency} & 2022 & IEEE ICDE\\
 \cite{zhouAHPAAdaptiveHorizontal2023} & Microservices & Cloud & Horizontal & {Heuristic\\~Transformer-based\\Queuing Theory} & Business Metrics & Hybrid & {Resource Efficiency\\Cost Efficiency\\SLA Assurance} & 2023 & AAAI \\
 \cite{shiautoscalingContainerizedApplications2023} & Microservices & Cloud & Hybrid & {Heuristic\\Deep Learning\\Reinforcement Learning} & Business Metrics & Proactive & {Cost Efficiency\\SLA Assurance} & 2023 & IEEE TSC\\
 \cite{chowDeepRestDeepResource2022} & Microservices & Cloud & Vertical & {Deep Learning\\~Transformer-based} & Hybrid & Proactive & {Resource Efficiency\\SLA Assurance} & 2022 & ACM EuroSys\\
 \cite{xueMetaReinforcementLearning2022} & Monolithic & Cloud & Horizontal & {Machine Learning\\Transformer-based\\Reinforcement Learning} & Business Metrics & Proactive & Resource Efficiency & 2022 & ACM SIGKDD\\
 \cite{chengProScaleProactiveAutoscaling2023} & Microservices & Edge & Horizontal & Heuristic & Business Metrics & Proactive & {Resource Efficiency\\SLA Assurance} & 2023 & IEEE TPDS\\
 \cite{xuEsDNNDeepNeural2022} & Microservices & Cloud & Horizontal & Deep Learning & System Metrics & Proactive & Resource Efficiency & 2022 & ACM TOIT\\
 \cite{fengHeterogeneityawareProactiveElastic2024} & Serverless & Cloud & Horizontal & {Deep Learning\\Heuristic} & Business Metrics & Proactive & {Cost Efficiency\\SLA Assurance} & 2024 & IEEE TSC\\
 \cite{liuScaleFluxEfficientStateful2022} & Microservices & Cloud & Vertical & \begin{tabular}[c]{@{}c@{}}Deep Learning\\Queuing Theory\end{tabular} & Business Metrics & Proactive & \begin{tabular}[c]{@{}c@{}}Resource Efficiency\\SLA Assurance \end{tabular} & 2022 & IEEE TPDS\\
   \cite{Wen2025} & Microservices & Cloud & Hybrid & \begin{tabular}[c]{@{}c@{}}Control Theory\\Learning-based\end{tabular} & System Metrics & Proactive & \begin{tabular}[c]{@{}c@{}}Cost Efficiency\end{tabular} & 2025 & ACM TAAS
\end{tblr}
}
\end{table*}

Predicting resource demand and performance requirements is a core component of behavior modeling in microservice auto-scaling. Many studies model the time series of request arrivals in containerized microservices, leveraging business metrics (reflecting user behavior) and machine metrics (capturing system and hardware status), as summarized in Table~\ref{Workload Characterization}.

Classical statistical methods were initially used for short- to medium-term forecasting of stationary workloads. For instance, Cheng et al. \cite{chengProScaleProactiveAutoscaling2023} applied simple moving average (SMA) in ProScale for edge workloads, while Qian et al. \cite{qianRobustScalerQoSAwareAutoscaling2022} used non-homogeneous poisson processes (NHPP) in RobustScaler. These approaches are limited in handling high-dimensional, nonlinear, and dynamic workloads, and often overlook instance-level resource contention.

Deep learning techniques improve modeling of complex patterns. Jeong et al. \cite{jeongProactiveResourceAutoscaling2023a} used SCINet for resource prediction, Liu et al. \cite{liuScaleFluxEfficientStateful2022} applied convolutional neural network (CNN) for bandwidth forecasting, and Xu et al. \cite{xuEsDNNDeepNeural2022} applied gated recurrent unit (GRU) for resource optimization. While these methods capture nonlinear trends, they may struggle with short-lived workloads, sudden spikes, long-term predictions, and control-plane overhead.

Ensemble approaches integrate multiple models to improve robustness. Feng et al. \cite{fengHeterogeneityawareProactiveElastic2024} combined machine learning and deep learning for elastic resource allocation, though centralized decision-making can limit scalability under high traffic. Wen et al. \cite{Wen2025} utilizes bothe control theory and learning-based approach make auto-scaling decisions.

Reinforcement learning based methods enhance proactive scaling by using workload predictions to guide scaling actions. Xu et al. \cite{xuCoScalMultifacetedScaling2022} and Shi et al. \cite{shiautoscalingContainerizedApplications2023} combined deep learning-based predictions with RL for adaptive scaling, improving resource utilization and SLA compliance. Challenges remain in managing resource contention, high-dimensional states, and implementation complexity in Kubernetes environments.

\begin{table*}
\centering
\caption{Classification Based on Performance Behavior.}
\label{Performance Analysis}
\resizebox{\linewidth}{!}{%
\begin{tblr}{
  cells = {c},
  hline{1,2,15} = {-}{0.08em},
  row{1-Z} = {m, 1cm},
}
\textbf{Reference} & \textbf{Architecture} & \textbf{Infrastructure} & \begin{tabular}[c]{@{}c@{}}\textbf{Scaling}\\\textbf{Method}\end{tabular} & \textbf{Technique} & \begin{tabular}[c]{@{}c@{}}\textbf{Scaling}\\\textbf{Indicator}\end{tabular} & \begin{tabular}[c]{@{}c@{}}\textbf{Scaling}\\\textbf{Timing}\end{tabular} & \textbf{Objective} & \textbf{Year} & \textbf{Source}\\ 
 \cite{lannurienHeROfakeHeterogeneousResources2023} & Serverless & Cloud & Horizontal & \begin{tabular}[c]{@{}c@{}}Heuristic\\Threshold-based\end{tabular} & Business Metrics & Reactive & \begin{tabular}[c]{@{}c@{}}Energy Efficiency\\SLA Assurance\end{tabular} & 2023 & ACM CCGrid\\ 
 \cite{chengGeoScaleMicroserviceAutoscaling2024} & Microservices & \begin{tabular}[c]{@{}c@{}}Cloud\\Edge\end{tabular} & Horizontal & \begin{tabular}[c]{@{}c@{}}Queuing Theory\\Heuristic\end{tabular} & Business Metrics & Proactive & \begin{tabular}[c]{@{}c@{}}Cost Efficiency\\SLA Assurance\end{tabular} & 2024 & IEEE TPDS\\ 
 \cite{rzadcaAutopilotWorkloadAutoscaling2020} & Microservices & Cloud & Hybrid & \begin{tabular}[c]{@{}c@{}}Machine Learning\\Heuristic\end{tabular} & System Metrics & Reactive & Resource Efficiency & 2020 & ACM EuroSys\\ 
 \cite{changBiCriteriaApproximationMultiOrigin2023} & Monolithic & Cloud & Horizontal & Heuristic & Business Metrics & Proactive & \begin{tabular}[c]{@{}c@{}}Cost Efficiency\\SLA Assurance\end{tabular} & 2023 & IEEE TMM\\ 
 \cite{caiInverseQueuingModelBased2022} & Microservices & \begin{tabular}[c]{@{}c@{}}Cloud \\Fog\\Edge\end{tabular} & Horizontal & \begin{tabular}[c]{@{}c@{}}Queuing Theory\\Control Theory\\Heuristic\end{tabular} & Business Metrics & Reactive & \begin{tabular}[c]{@{}c@{}}SLA Assurance\\Resource Efficiency\end{tabular} & 2022 & IEEE TC\\ 
 \cite{heidariCostEfficientautoscalingAlgorithm2021} & Monolithic & Cloud & Horizontal & \begin{tabular}[c]{@{}c@{}}Heuristic\\Threshold-based\end{tabular} & System Metrics & Reactive & \begin{tabular}[c]{@{}c@{}}Cost Efficiency\\Resource Efficiency\end{tabular} & 2021 & IEEE TSE\\ 
 \cite{baarziSHOWARRightSizingEfficient2021} & Microservices & Cloud & Hybrid & \begin{tabular}[c]{@{}c@{}}Heuristic\\Control Theory\end{tabular} & Hybrid & Reactive & \begin{tabular}[c]{@{}c@{}}Resource Efficiency\\SLA Assurance\end{tabular} & 2021 & ACM SoCC\\ 
 \cite{zhangElasticTaskOffloading2024} & Microservices & Cloud & Muti-faceted & \begin{tabular}[c]{@{}c@{}}Reinforcement Learning\\Deep Learning\\Queuing Theory\end{tabular} & Hybrid & Proactive & \begin{tabular}[c]{@{}c@{}}Resource Efficiency\\Energy Efficiency\\~Cost Efficiency\end{tabular} & 2024 & IEEE TNSM\\ 
 \cite{wangElasticSchedulingMicroservice2021} & Microservices & Cloud & Horizontal & Heuristic & System Metrics & Reactive & \begin{tabular}[c]{@{}c@{}}Cost Efficiency\\Resource Efficiency\end{tabular} & 2021 & IEEE TPDS\\ 
 \cite{liuCoordinatingFastConcurrency2022} & \begin{tabular}[c]{@{}c@{}}Monolithic\\Microservices\end{tabular} & Cloud & Horizontal & Control Theory & Hybrid & Reactive & SLA Assurance & 2022 & IEEE TPDS\\ 
 \cite{rossiDynamicMultiMetricThresholds2023} & Microservices & Cloud & Horizontal & \begin{tabular}[c]{@{}c@{}}Reinforcement Learning\\Threshold-based\end{tabular} & System Metrics & Reactive & \begin{tabular}[c]{@{}c@{}}SLA Assurance\\Cost Efficiency\end{tabular} & 2023 & IEEE TCC\\ 
 \cite{amiriCostAwareMultifacetedReconfiguration2023} & Monolithic & Cloud & Horizontal & \begin{tabular}[c]{@{}c@{}}Queuing Theory\\Heuristic\\Threshold-based\end{tabular} & Business Metrics & Reactive & \begin{tabular}[c]{@{}c@{}}Cost Efficiency\\Resource Efficiency\end{tabular} & 2023 & IEEE CLOUD\\
 \cite{SHAFI2025104266} & Microservices & Cloud & Horizontal & \begin{tabular}[c]{@{}c@{}}Machine\\Learning\end{tabular} & System Metrics & Reactive & \begin{tabular}[c]{@{}c@{}}Cost Efficiency\end{tabular} & 2025 & JNCA\\
\end{tblr}
}
\end{table*}

Transformer-based models have recently emerged for time series forecasting. Wen et al. \cite{wenTempoScaleCloudWorkloads2024} proposed TempoScale using long- and short-term features to optimize elasticity, Zhou et al. \cite{zhouAHPAAdaptiveHorizontal2023} applied time series decomposition with queuing theory, Chow et al. \cite{chowDeepRestDeepResource2022} leveraged deep learning for API-based resource estimation, and Xue et al. \cite{xueMetaReinforcementLearning2022} used meta-RL for improved allocation accuracy. While promising, these approaches require further refinement to handle microservice dependencies, real-time adaptation, and limited historical data.

\textit{\textbf{Observations:}} Classical methods are effective for short-term forecasts but struggle with dynamic workloads. Deep learning and ensemble methods better handle nonlinear and high-dimensional data but face challenges with scalability and sudden load changes. Reinforcement learning enhances adaptive scaling but introduces complexity, while Transformer-based models show potential for long-term dependencies but need optimization for microservice-specific challenges. Overall, workload characterization and prediction remain active areas with significant scope for innovation in microservice auto-scaling.

\subsubsection{\textbf{Performance Behavior}}

Performance behavior in microservice auto-scaling provides real-time insights to guide dynamic resource adjustments, ensuring system efficiency, stability, and SLA compliance. Table~\ref{Performance Analysis} summarizes recent studies leveraging this technique.

SLA-driven approaches focus on maintaining expected performance and user experience. Chang and Chan \cite{changBiCriteriaApproximationMultiOrigin2023} proposed a bi-criteria approximation algorithm to reduce deployment costs while satisfying QoS constraints, though with high computational complexity for large-scale systems. Cheng et al. \cite{chengGeoScaleMicroserviceAutoscaling2024} applied Lyapunov optimization for distributed edge clouds, but did not fully address heterogeneous processing rates. Lannurien et al. \cite{lannurienHeROfakeHeterogeneousResources2023} optimized scaling across CPU, GPU, and FPGA nodes, improving response time and energy efficiency, yet practical deployment challenges remain. Control- and queuing theory-based methods, including Baarzi and Kesidis \cite{baarziSHOWARRightSizingEfficient2021}, Liu et al. \cite{liuCoordinatingFastConcurrency2022}, and Cai and Buyya \cite{caiInverseQueuingModelBased2022}, enhance resource allocation and stability but often lack predictive capability or incur overhead under dynamic workloads. Multi-agent reinforcement learning \cite{rossiDynamicMultiMetricThresholds2023} allows adaptive threshold tuning but introduces training complexity and scalability limits.

%Cheng et al. They decomposed the long-term optimization problem into a series of single-slot sub-problems using Lyapunov optimization, and proposed SGP algorithm to solve them in an online manner for near-optimal solutions by adaptively resizing microservice instances at geographically distributed edge clouds. Nevertheless, the GeoScale solution does not consider automatic scaling of microservices in heterogeneous edge clouds (since instances of the same template service are likely to have different processing rates when deployed across a variety edge cloud)

\begin{table*}
\centering
\caption{Classification Based on Anomaly Awareness.}
\label{Anomaly Detection}
\resizebox{\linewidth}{!}{
\begin{tblr}{
  cells = {c},
  hline{1,2,13} = {-}{0.08em},
  row{1-Z} = {m, 1cm},
}
\textbf{Reference} & \textbf{Architecture} & \textbf{Infrastructure} & \begin{tabular}[c]{@{}c@{}}\textbf{Scaling}\\\textbf{Method}\end{tabular} & \textbf{Technique} & \begin{tabular}[c]{@{}c@{}}\textbf{Scaling}\\\textbf{Indicator}\end{tabular} & \begin{tabular}[c]{@{}c@{}}\textbf{Scaling}\\\textbf{Timing}\end{tabular} & \textbf{Objective}  & \textbf{Year} & \textbf{Source}\\ 
 \cite{I19} & Microservices & Cloud & Hybrid & \begin{tabular}[c]{@{}c@{}}Machine Learning\\Reinforcement Learning\end{tabular} & Hybrid & Proactive & SLA Assurance & 2020 & USENIX OSDI\\ 
 \cite{caiAutoManResourceefficientProvisioning2023} & Microservices & Cloud & Vertical & Reinforcement Learning & Hybrid & Proactive & \begin{tabular}[c]{@{}c@{}}Resource Efficiency\\SLA Assurance\end{tabular}  & 2023 & Future Gener. Comput. Syst.\\ 
 \cite{ganSagePracticalScalable2021} & Microservices & Cloud & Hybrid & Deep Learning & Hybrid & Proactive & \begin{tabular}[c]{@{}c@{}}Resource Efficiency\\SLA Assurance\end{tabular}  & 2021 & ACM ASPLOS\\ 
 \cite{abdullahBurstAwarePredictiveautoscaling2022} & Microservices & Cloud & Horizontal & \begin{tabular}[c]{@{}c@{}}Machine Learning\\Threshold-based\end{tabular} & Business Metrics & Hybrid & \begin{tabular}[c]{@{}c@{}}SLA Assurance\\Cost Efficiency\\Resource Efficiency\end{tabular}  & 2022 & IEEE TSC\\ 
 \cite{zhangLearningdrivenHybridScaling2024} & Microservices & Cloud & Hybrid & \begin{tabular}[c]{@{}c@{}}Deep Learning\\Reinforcement Learning\\Threshold-based\end{tabular} & Hybrid & Hybrid & \begin{tabular}[c]{@{}c@{}}Resource Efficiency\\SLA Assurance\end{tabular}  & 2024 & J. Parallel Distrib. Comput.\\ 
 \cite{sachidanandaErlangApplicationAwareautoscaling2024} & Microservices & Cloud & Horizontal & \begin{tabular}[c]{@{}c@{}}Queuing Theory\\Machine Learning\end{tabular} & Business Metrics & Hybrid & Cost Efficiency  & 2024 & ACM EuroSys\\ 
 \cite{qiuAWAREAutomateWorkload} & Microservices & Cloud & Hybrid & \begin{tabular}[c]{@{}c@{}}Machine Learning\\Reinforcement Learning\end{tabular} & Hybrid & Proactive & SLA Assurance & 2023 & USENIX ATC\\ 
 \cite{liuMConAdapterReinforcementLearningbased2023} & Microservices & Cloud & Hybrid & Reinforcement Learning & Business Metrics & Proactive & \begin{tabular}[c]{@{}c@{}}Resource Efficiency\\Cost Efficiency\\SLA Assurance\end{tabular}  & 2023 & ACM SoCC\\ 
 \cite{zhuQoSAwareCoSchedulingDistributed2022} & Microservices & Cloud & Vertical & \begin{tabular}[c]{@{}c@{}}Machine Learning\\Heuristic\end{tabular} & Business Metrics & Reactive & \begin{tabular}[c]{@{}c@{}}Resource Efficiency\\SLA Assurance\end{tabular}  & 2022 & IEEE TPDS\\ 
 \cite{xiePBScalerBottleneckAwareautoscaling2024} & Microservices & Cloud & Horizontal & Machine Learning & Hybrid & Proactive & \begin{tabular}[c]{@{}c@{}}Resource Efficiency\\SLA Assurance\end{tabular}  & 2024 & IEEE TSC\\
 \cite{shiNodensEnablingResource} & Microservices & Cloud & Vertical & {Machine Learning\\Queuing Theory} & System Metrics & Reactive & {Resource Efficiency\\SLA Assurance} & 2023 & USENIX ATC
\end{tblr}
}
\end{table*}

Resource efficiency approaches optimize allocation to reduce cost and utilization. Heidari and Buyya \cite{heidariCostEfficientautoscalingAlgorithm2021} addressed performance-cost trade-offs in large-scale graph processing, and Wang et al. \cite{wangElasticSchedulingMicroservice2021} proposed elastic scheduling for microservices, though both require adaptation for heterogeneous, dynamic environments. Reinforcement learning-based methods, such as Rzadca et al. \cite{rzadcaAutopilotWorkloadAutoscaling2020}, Zhang et al. \cite{zhangElasticTaskOffloading2024}, Amiri and Zdun \cite{amiriCostAwareMultifacetedReconfiguration2023}, and Shafi \cite{SHAFI2025104266}, improve adaptive scaling and resource efficiency but demand significant computation, historical data, and careful parameter tuning.

\textit{\textbf{Observations:}} Performance analysis in microservice auto-scaling primarily targets SLA assurance and resource efficiency. Classical heuristics, control theory, queuing theory, optimization algorithms, and reinforcement learning enhance allocation and stability but face challenges in computational overhead, dynamic adaptability, and real-time response under heterogeneous workloads. Recent approaches show promise but require further empirical validation and refinement to handle complex microservice dependencies effectively.

\begin{table*}
\centering
\caption{Classification Based on Dependency Modeling.}
\label{Dependency Analysis}
\resizebox{\linewidth}{!}{
\begin{tblr}{
  cells = {c},
  hline{1,2,12} = {-}{0.08em},
  row{1-Z} = {m, 1cm},
}
\textbf{Reference} & \textbf{Architecture} & \textbf{Infrastructure} & \begin{tabular}[c]{@{}c@{}}\textbf{Scaling}\\\textbf{Method}\end{tabular} & \textbf{Technique} &  \begin{tabular}[c]{@{}c@{}}\textbf{Scaling}\\\textbf{Indicator}\end{tabular} & \begin{tabular}[c]{@{}c@{}}\textbf{Scaling}\\\textbf{Timing}\end{tabular} & \textbf{Objective}  & \textbf{Year} & \textbf{Source}\\ 
 \cite{mengDeepScalerHolisticAutoscaling2023a} & Microservices & Cloud & Horizontal & \begin{tabular}[c]{@{}c@{}}Deep Learning\\Threshold-based\end{tabular} & Business Metrics & Proactive & \begin{tabular}[c]{@{}c@{}}Resource Efficiency\\SLA Assurance\end{tabular}  & 2023 & ACM ASE\\ 
 \cite{zhangSinanMLbasedQoSaware2021} & Microservices & Cloud & Vertical & \begin{tabular}[c]{@{}c@{}}Machine Learning\\Deep Learning\end{tabular} & Hybrid & Proactive & \begin{tabular}[c]{@{}c@{}}Resource Efficiency\\SLA Assurance\end{tabular}  & 2021 & ACM ASPLOS\\ 
 \cite{zengTopologyAwareSelfAdaptiveResource2023} & Microservices & Cloud & Vertical & Reinforcement Learning & Hybrid & Proactive & \begin{tabular}[c]{@{}c@{}}Resource Efficiency\\SLA Assurance\end{tabular}  & 2023 & IEEE ICWS\\ 
 \cite{hossenPracticalEfficientMicroservice2022} & Microservices & Cloud & Hybrid & Control Theory & Hybrid & Proactive & \begin{tabular}[c]{@{}c@{}}Resource Efficiency\\SLA Assurance\end{tabular}  & 2022 & ACM HPDC\\ 
 \cite{mirhosseiniParsloGradientDescentbased2021} & Microservices & Cloud & Horizontal & \begin{tabular}[c]{@{}c@{}}Machine Learning\\Heuristic\end{tabular} & Business Metrics & Proactive & \begin{tabular}[c]{@{}c@{}}Resource Efficiency\\Cost Efficiency\\SLA Assurance\end{tabular}  & 2021 & ACM SoCC\\ 
 \cite{tongGMAGraphMultiagent2023} & Microservices & \begin{tabular}[c]{@{}c@{}}Edge \\Cloud\end{tabular} & Horizontal & \begin{tabular}[c]{@{}c@{}}Deep Learning\\Reinforcement Learning\end{tabular} & Hybrid & Proactive & SLA Assurance  & 2023 & IEEE ICWS\\ 
 \cite{songChainsFormerChainLatencyAware2023} & Microservices & Cloud & Hybrid & \begin{tabular}[c]{@{}c@{}}Machine Learning\\Reinforcement Learning\end{tabular} & System Metrics & Proactive &  \begin{tabular}[c]{@{}c@{}}Resource Efficiency\\SLA Assurance\end{tabular}  & 2023 & Springer SOC\\ 
 \cite{wangDeepScalingMicroservicesautoscaling2022} & Microservices & Cloud & Hybrid & \begin{tabular}[c]{@{}c@{}}Deep Learning\\Reinforcement Learning\end{tabular} & Hybrid & Proactive & \begin{tabular}[c]{@{}c@{}}Resource Efficiency\\Cost Efficiency\\SLA Assurance\end{tabular}  & 2022 & ACM SoCC\\
  \cite{LIANG2026107909} & Microservices & Cloud & Vertical & \begin{tabular}[c]{@{}c@{}}Deep Learning\end{tabular} & System Metrics & Proactive & \begin{tabular}[c]{@{}c@{}}Resource Efficiency\end{tabular}  & 2025 & FGCS\\
    \cite{Fang2025} & Microservices & Cloud & Hybrid & \begin{tabular}[c]{@{}c@{}}Graph Neural\\ Network\end{tabular} & System Metrics & Proactive & \begin{tabular}[c]{@{}c@{}}Resource Efficiency\end{tabular}  & 2025 & IEEE TSC\\
\end{tblr}
}
\end{table*}
\subsubsection{\textbf{Anomaly Awareness}}

Monitoring and detecting abnormal behaviors in microservices enables identification of potential bottlenecks, which often constrain overall system performance according to the Law of the Minimum. Table~\ref{Anomaly Detection} presents recent studies employing this technique.

Most approaches leverage machine learning and deep learning. Xie et al. \cite{xiePBScalerBottleneckAwareautoscaling2024} combined TopoRank and genetic algorithms to reduce resource consumption and maintain performance, but bottleneck detection can be slow, delaying scaling. Shi et al. \cite{shiNodensEnablingResource} used linear regression for real-time allocation, though linear models struggle under high load variations. Zhu et al. \cite{zhuQoSAwareCoSchedulingDistributed2022} applied predictive allocation for microservices, yet scalability and manual tuning remain issues. Sachidananda and Sivaraman \cite{sachidanandaErlangApplicationAwareautoscaling2024} used multi-armed bandits for VM allocation, but training is time-consuming. Abdullah et al. \cite{abdullahBurstAwarePredictiveautoscaling2022} proposed burst-aware scaling, which may fail for irregular small bursts. Graph-based methods like Gan et al. \cite{ganSagePracticalScalable2021} improve performance issue localization but struggle with unseen issues.

Reinforcement learning is increasingly applied to mitigate bottlenecks. Qiu et al. \cite{I19} combined SVM and deep deterministic policy gradient (DDPG) for SLA-aware scaling, though early training can cause instability. Cai et al. \cite{caiAutoManResourceefficientProvisioning2023} used multi-agent DDPG for large-scale resource management, but with high communication overhead. Meta-learning and DRL based approaches \cite{qiuAWAREAutomateWorkload, liuMConAdapterReinforcementLearningbased2023} support adaptive scaling, yet face slow adaptation or limited resource types. Zhang et al. \cite{zhangLearningdrivenHybridScaling2024} applied DRL for hybrid scaling, improving allocation accuracy but requiring extensive data and long training times.

\textit{\textbf{Observations:}} Current anomaly-awareness methods advance bottleneck detection and resource efficiency in microservice environments but face challenges in computational overhead, training cost, dynamic adaptation, and data requirements. Reinforcement learning shows promise for adaptive and predictive scaling, yet improving stability, scalability, and generality remains an open research direction.

\subsubsection{\textbf{Dependency Modeling}}

Dependency modeling analyzes the interactions between microservices, enabling optimized communication and ensuring system scalability and stability. Table~\ref{Dependency Analysis} presents recent studies falling into this classification.

Early methods relied on simple feedback or gradient-based techniques. Hossen et al. \cite{hossenPracticalEfficientMicroservice2022} used feedback-based resource management, but it struggled with complex service dependencies. Mirhosseini et al. \cite{mirhosseiniParsloGradientDescentbased2021} applied Parslo gradient descent to reduce deployment costs while meeting SLAs, yet response times lagged under rapid traffic changes. Zhang et al. \cite{zhangSinanMLbasedQoSaware2021} employed CNN and Boosted Trees to model inter-service dependencies, improving latency compliance, but at high computational cost.

%As application demands and system complexity increase, simple methods face challenges. Some studies use reinforcement learning to improve decision-making. Song et al. \cite{songChainsFormerChainLatencyAware2023} introduced the ChainsFormer framework, which combines machine learning and reinforcement learning to optimize scaling decisions, but it requires significant time for initial training and model updates, and its performance under extreme loads is unverified. Wang et al. \cite{wangDeepScalingMicroservicesautoscaling2022} improved load prediction and CPU utilization estimation with deep learning and reinforcement learning, but it demands substantial computational resources, making it less suitable for small-scale systems.

\begin{table*}
\centering
\caption{Classification Based on Co-location Interference.}
\label{Task Co-location}
\resizebox{\linewidth}{!}{
\begin{tblr}{
  cells = {c},
  hline{1,2,6} = {-}{0.08em},
  row{1-Z} = {m, 1cm},
}
\textbf{Reference} & \textbf{Architecture} & \textbf{Infrastructure} & \begin{tabular}[c]{@{}c@{}}\textbf{Scaling}\\\textbf{Method}\end{tabular} & \textbf{Technique} & \begin{tabular}[c]{@{}c@{}}\textbf{Scaling}\\\textbf{Indicator}\end{tabular} & \begin{tabular}[c]{@{}c@{}}\textbf{Scaling}\\\textbf{Timing}\end{tabular} & \textbf{Objective}  & \textbf{Year} & \textbf{Source}\\  
\cite{luoErmsEfficientResource2022} & Microservices & Cloud & Horizontal & \begin{tabular}[c]{@{}c@{}}Queuing Theory\\Heuristic\end{tabular} & Business Metrics & Reactive & \begin{tabular}[c]{@{}c@{}}SLA Assurance\\Resource Efficiency\end{tabular}  & 2022  & ACM ASPLOS\\
\cite{10501918} & Microservices & Cloud & Horizontal & \begin{tabular}[c]{@{}c@{}}Machine Learning\\Threshold-based\end{tabular} & Business Metrics & Proactive & \begin{tabular}[c]{@{}c@{}}SLA Assurance\\Resource Efficiency\end{tabular} & 2023  & IEEE SmartData\\
\cite{1021007} & Microservices & Cloud & Horizontal & Heuristic & System Metrics & Proactive & Resource Efficiency & 2020  & J. Comput. Sci. Technol.\\
\cite{9242282} & Microservices & Cloud & Horizontal & \begin{tabular}[c]{@{}c@{}}Queuing Theory\\Control Theory\end{tabular} & System Metrics & Proactive & \begin{tabular}[c]{@{}c@{}}SLA Assurance\\Resource Efficiency\end{tabular} & 2022 & IEEE TCC
\end{tblr} 
}
\end{table*}

As systems grew more complex, reinforcement learning and deep learning were adopted. Song et al. \cite{songChainsFormerChainLatencyAware2023} combined ML and RL for scaling decisions, though initial training and extreme load performance remain concerns. Wang et al. \cite{wangDeepScalingMicroservicesautoscaling2022} enhanced load prediction with deep learning and RL, but required significant resources. To capture intricate service relationships, GNN-based methods emerged: Meng et al. \cite{mengDeepScalerHolisticAutoscaling2023a} used spatiotemporal GNNs with adaptive graph learning, Zeng et al. \cite{zengTopologyAwareSelfAdaptiveResource2023} combined GNNs with RL for resource allocation, and Tong et al. \cite{tongGMAGraphMultiagent2023} applied GNNs in edge clouds. Fang et al \cite{Fang2025} proposed a hierarchical GNN to capture dependencies in container-based clouds. These approaches improved SLA compliance and resource efficiency but still face challenges under high concurrency, limited resources, or poor historical data.

\textit{\textbf{Observations:}} As microservice systems become more complex, dependency modeling requires advanced methods such as GNNs and RL to accurately capture interactions. Despite progress, challenges remain in computational cost, scalability, and handling dynamic, high-concurrency workloads, leaving substantial room for further research.

\subsubsection{\textbf{Co-location Interference}}

Task co-location optimizes the placement of multiple microservice tasks on shared physical resources to maximize utilization and system performance while maintaining SLA guarantees. Table~\ref{Task Co-location} presents recent studies in this category.

Li et al. \cite{10501918} introduced a machine learning-based scheduling approach using predicted disturbances to minimize initial scheduling latency but did not address dynamic adjustments. Chen et al. \cite{1021007} analyzed hardware events under two co-location scenarios, providing deployment guidance, but lacked coverage of more complex configurations. Jiang et al. \cite{9242282} studied co-location of online and batch jobs in Alibaba Cloud, uncovering utilization bottlenecks, though results were dataset-specific. Luo et al. \cite{luoErmsEfficientResource2022} proposed Erms, an efficient resource management system that improves utilization and reduces SLA violations through dependency-aware scheduling, yet its complexity and maintenance costs remain high.

\textit{\textbf{Observations:}} Task co-location enhances resource efficiency, performance, and cost-effectiveness, but increasing microservice complexity and dynamic workloads present ongoing challenges. Future work should focus on adaptive, lightweight algorithms that balance precision, scalability, and operational overhead for diverse cloud environments.

\section{Challenges and Future Directions}\label{5}

Despite extensive research on auto-scaling for microservice applications, several challenges remain. Here, we highlight key research opportunities and directions for future work:
\begin{itemize} 
    
\item [$\bullet$] \textbf{Balancing Model Complexity and Overhead:} Many studies employ complex models to improve prediction accuracy, but these incur significant computational and operational overhead. Lightweight models often provide sufficient accuracy and are more practical for resource-constrained environments. Future work should focus on methods that balance accuracy, interpretability, and efficiency, reserving complex models for scenarios where high precision is critical.

\item [$\bullet$] \textbf{Modeling Microservice Dependencies:} Interactions between microservices mean that scaling one service can impact others. Auto-scaling strategies should leverage dependency graphs and monitor metrics across the service call chain, coordinating scaling decisions to avoid bottlenecks and ensure system-wide stability. Predictive methods for load transfer between services can further improve resource allocation.

\item [$\bullet$] \textbf{Leveraging Large Models for Generalization:} Microservice workloads vary widely across domains (e.g., e-commerce, streaming). Traditional models often struggle to generalize across heterogeneous load patterns. Incorporating large models, such as Transformer-based architectures or GPT-like models, can capture diverse patterns and facilitate transfer learning to adapt rapidly to new environments.

\item [$\bullet$] \textbf{Multidimensional Performance Evaluation:} Effective auto-scaling requires monitoring multiple metrics, including CPU, memory, storage, latency, throughput, and error rates. Holistic evaluation enables identification of bottlenecks, SLA compliance, and optimization of resource allocation. Future research should develop adaptive, multidimensional monitoring frameworks to guide smarter scaling decisions across complex microservice systems.

%\item [$ \bullet $] \textbf{Explore Multi-faceted Scaling:} Multi-faceted scaling, including techniques like brownout, can optimize resource utilization and improve system stability. While not widely adopted due to complexity, these methods have great potential to improve performance in container environments. As technology advances, more companies may start adopting these methods to improve efficiency.

 \item [$ \bullet $] \textbf{Enhance Adaptability with Meta-learning:} Meta-learning enables auto-scaling systems to quickly adapt to new workloads and environments by updating scaling strategies in real time based on observed system feedback \cite{F5}. This approach improves responsiveness and maintains optimal performance under varying conditions.

%\item [$ \bullet $] \textbf{Consider Performance Interference Between Applications:} In multi-tenant environments, cloud-native applications share resources like CPU, memory, and disk I/O, which can lead to performance interference. Solutions include refining resource isolation, optimizing resource allocation algorithms, and using cross-layer optimization strategies to reduce conflicts. Adaptive load-balancing technologies can also help distribute load evenly across applications and servers, minimizing interference.
\end{itemize} 

\section{Summary and Conclusions}\label{6}

In this study, we conducted a comprehensive review of auto-scaling technologies for resource management in microservice applications. Using the taxonomy proposed in Section~\ref{3}, we systematically categorized and analyzed existing approaches, emphasizing behavior modeling and objective optimization. This organization allowed us to highlight key trends, evaluate the effectiveness of various methods, and assess their applicability across different microservice scenarios. Our analysis revealed the limitations of current approaches, particularly in handling complex dependencies, dynamic workloads, and resource contention. Based on these insights, we outlined several promising research directions, encouraging the development of smarter, more adaptive, and efficient auto-scaling strategies to advance resource management in microservice systems.

\bibliographystyle{unsrt1}
\bibliography{sample-base}

\begin{IEEEbiography}[{\includegraphics[width=1in,height=1.2in,clip,keepaspectratio]{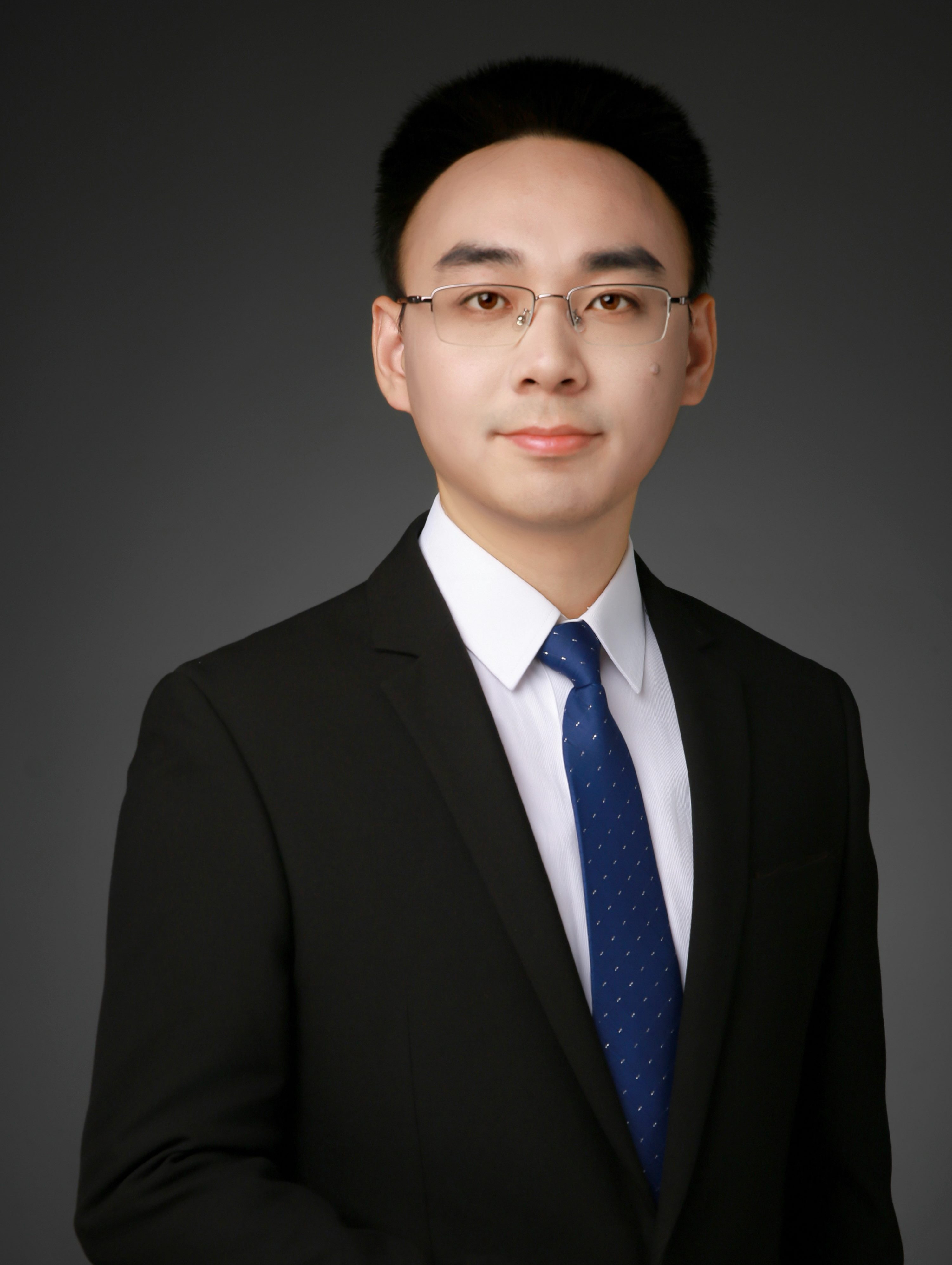}}]{Minxian Xu}
(Senior Member, IEEE) is currently an Associate Professor at the Shenzhen Institutes of Advanced Technology, Chinese Academy of Sciences. He received his PhD degree from the University of Melbourne in 2019. His research interests include resource management for cloud-native cluster and applications. He has co-authored over 80 peer-reviewed papers published in prominent international journals and conferences with 6000+ citations. He was awarded the 2023 IEEE TCSC Early Career Award (for contributions in efficient management of large-scale microservice-based cluster). He is among the world's top 2\% scientists by Stanford University in 2023 and 2024. He is also the senior member of IEEE and CCF.
\end{IEEEbiography}

\begin{IEEEbiography}[{\includegraphics[width=1in,height=1.2in,clip,keepaspectratio]{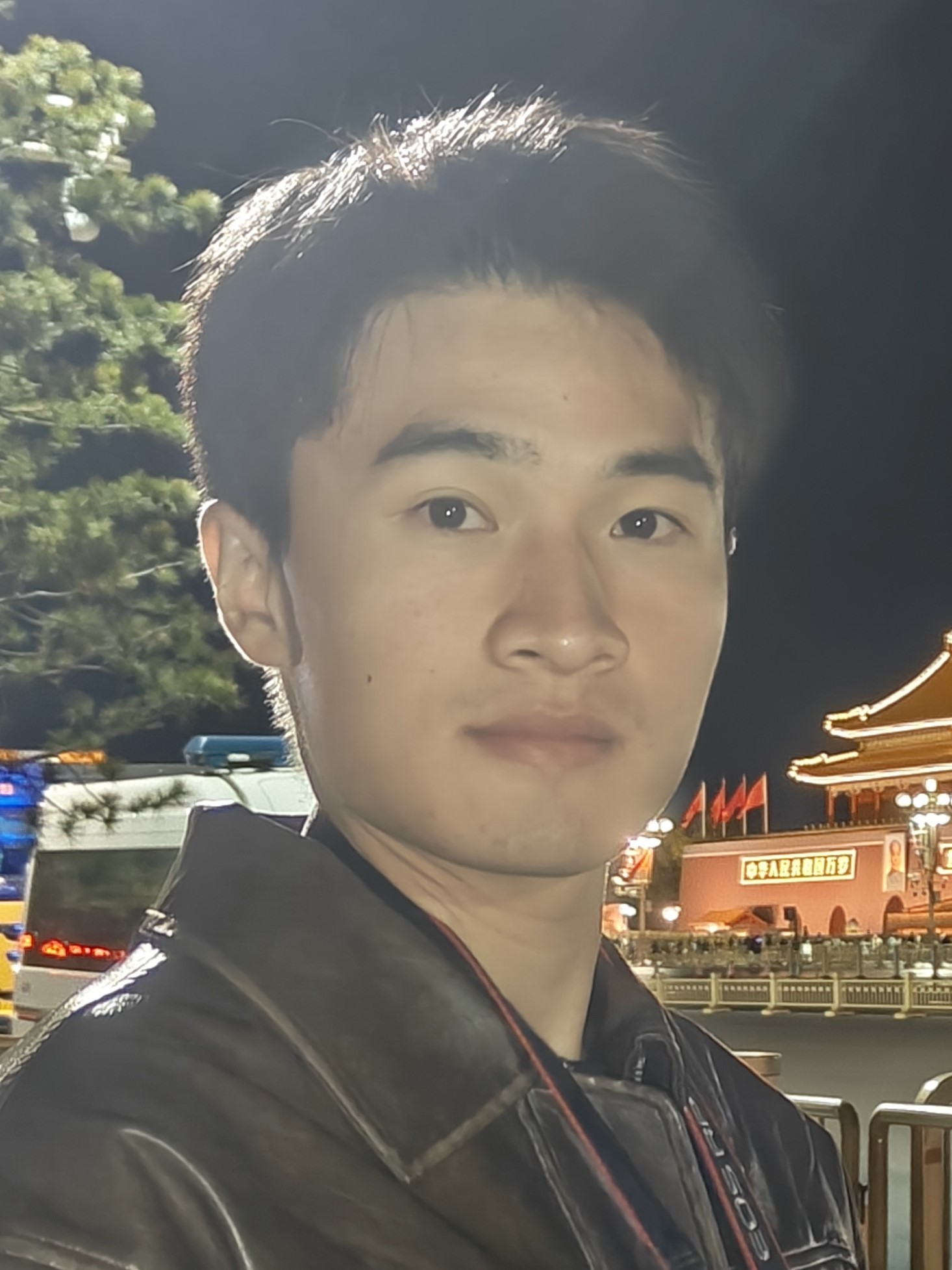}}]{Junhan Liao}
 received his BSc degree from the Hunan University of Technology. Now he is a master student at the University of the Chinese Academy of Sciences. He conducts scientific research under the guidance of his advisor at the Shenzhen Institutes of Advanced Technology, Chinese Academy of Sciences. His primary research focuses on the characterization of inference workload features and inference optimization in large language models.
\end{IEEEbiography}

\begin{IEEEbiography}[{\includegraphics[width=1in,height=1.2in,clip,keepaspectratio]{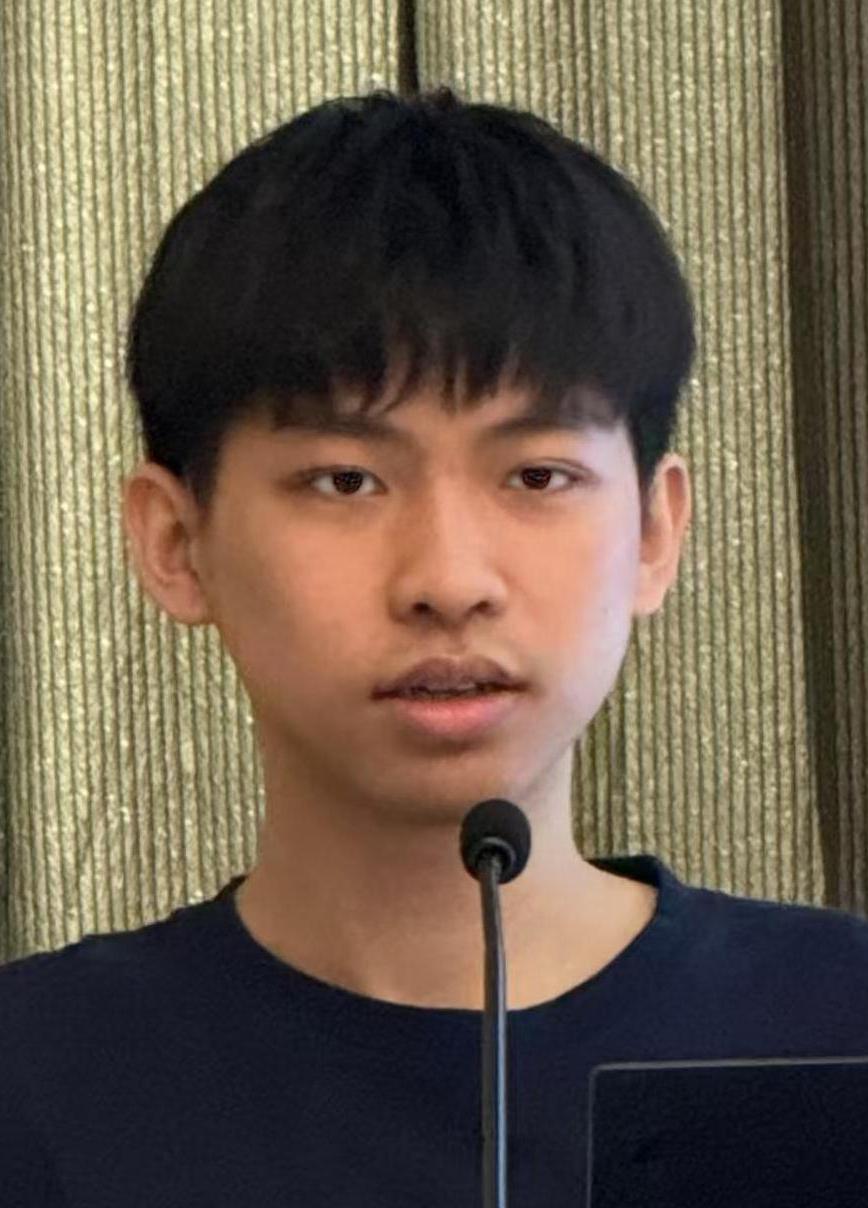}}]{Linfeng Wen}
 received his BSc degree from the Guangdong Ocean University. Now he is a master student at the University of the Chinese Academy of Sciences, and conducting research at Shenzhen Institutes of Advanced Technology, Chinese Academy of Sciences. His primary research focuses on the characterization of workload features and resource management in cloud-native applications. He has published several papers at ACM TAAS, SPE, IEEE ISPA and IEEE CLOUD.
\end{IEEEbiography}

%\vspace{-2cm}

\begin{IEEEbiography}[{\includegraphics[width=1in,height=1.2in,clip,keepaspectratio]{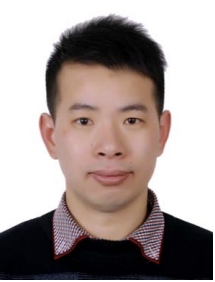}}]{Huaming Wu}
 (Senior Member, IEEE) received the BE and MS degrees from the Harbin Institute of Technology, China, in 2009 and 2011, respectively, both in electrical engineering, and the PhD degree in the highest honor in computer science from Freie Universität Berlin, Germany, in 2015. He is currently
a professor at the Center for Applied Mathematics, Tianjin University, China. His research interests include mobile cloud computing, edge computing, Internet of Things, and DNA storage.
\end{IEEEbiography}

%\vspace{-2cm}

\begin{IEEEbiography}[{\includegraphics[width=1in,height=1.2in,clip,keepaspectratio]{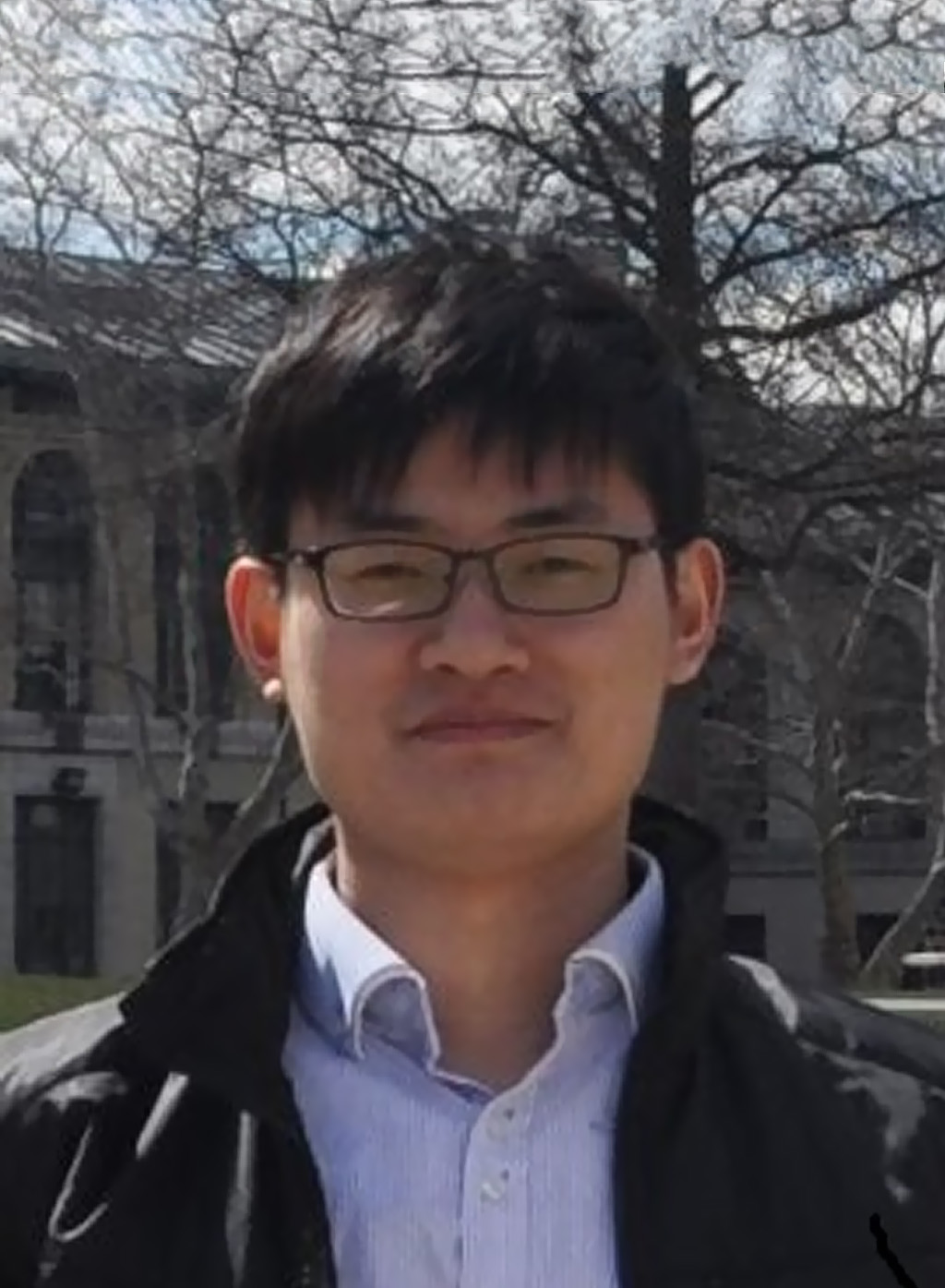}}]{Kejiang Ye}
(Senior Member, IEEE) received the BSc and PhD degrees from Zhejiang University in 2008 and 2013, respectively. He was also a joint PhD student with the University of Sydney from 2012 to 2013. After graduation, he worked as a postdoctoral researcher at Carnegie Mellon University from 2014 to 2015 and at Wayne State University from 2015 to 2016. He is currently a professor at the Shenzhen Institutes of Advanced Technology, Chinese Academy of Sciences. His research interests focus on the performance, energy, and reliability of cloud computing and network.
\end{IEEEbiography}

%\vspace{-2cm}

\begin{IEEEbiography}[{\includegraphics[width=1in,height=1.2in,clip,keepaspectratio]{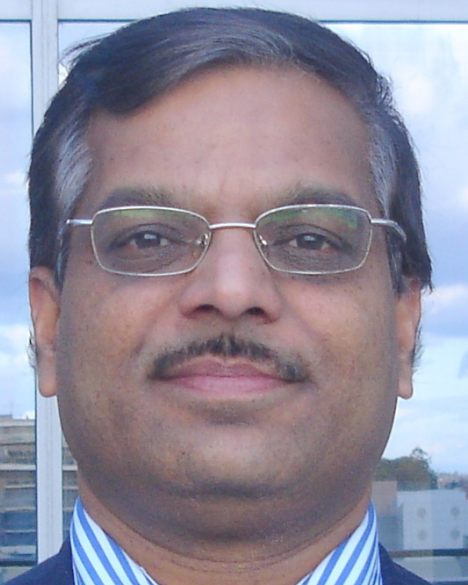}}]{Rajkumar Buyya}
(Fellow, IEEE) is a Redmond Barry Distinguished Professor and Director of the Quantum Cloud Computing and Distributed Systems (qCLOUDS) Laboratory at the University of Melbourne, Australia. He has authored more than 850 publications and seven textbooks, including Mastering Cloud Computing published by McGraw-Hill, China Machine Press, and Morgan Kaufmann. Recognized as one of the world’s most highly cited researchers in computer science and software engineering (h-index: 178; g-index: 394; 168,800+ citations), Dr. Buyya is a Fellow of IEEE, a Foreign Fellow of Academia Europaea, and a Fellow of ACM. He co-founded five major IEEE/ACM international conferences—CCGrid, Cluster, Grid, e-Science, and UCC—and served as the Chair of their inaugural meetings. He served as the founding Editor-in-Chief of the IEEE Transactions on Cloud Computing. He is currently serving as Co-Editor-in-Chief of Journal of Software: Practice and Experience, which was established 55+ years ago.
\end{IEEEbiography}

\vspace{-12cm}

\begin{IEEEbiography}[{\includegraphics[width=1in,height=1.2in,clip,keepaspectratio]{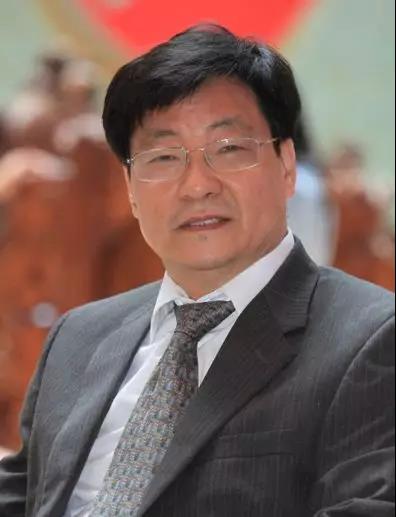}}]{Chengzhong Xu}
(Fellow, IEEE) received the Ph.D. degree in Computer Science and Engineering from the University of Hong Kong in 1993. He is the Dean of the Faculty of Science and Technology and the Interim Director of the Institute of Collaborative Innovation at the University of Macau. Dr. Xu’s research focuses on parallel and distributed computing, with an emphasis on resource management for performance, reliability, availability, power efficiency, and security. His work spans servers and cloud datacenters,  wireless embedded devices and edge AI systems , with applications in smart city and autonomous driving. He has authored two research monographs and more than 600 papers, which have garnered more than 25K citations with an H-index of 82, and has been cited in over 300 international patents (including 230 USA patents as of year 2024, per SciVal).
\end{IEEEbiography}

\end{document}